%% file: combesf.tex
\documentclass[multphys,vecphys]{svmult}

\usepackage{makeidx}     
\usepackage{graphicx}    
\usepackage{multicol}    

\makeindex             

\begin{document}

\title*{CO emission associated with a cooling flow}
\author{F. Combes\inst{}\and
P. Salom\'e\inst{}}
\institute{LERMA, Observatoire de Paris, 61 Av. de l'Observatoire,
75014, Paris, France
\texttt{francoise.combes@obspm.fr,}
\texttt{philippe.salome@obspm.fr}}
%
%
\maketitle

\section{Abstract}
The existence of cooling flows in the center of galaxy clusters has always been
a puzzle, and in particular the fate of the cooling gas, since the presence of
cold gas has never been proven directly.
X-ray data from the satellites Chandra and XMM-Newton have  constrained
the amount of cooling, and it was realized that feedback and heating
from a central AGN and its jets could reduce the amount of cold gas.
 Recently, a few central galaxies of cooling flow clusters
have been detected in the CO lines. 
For the first time, we show IRAM interferometer maps of CO(1-0) and CO(2-1)
emission in a cooling flow,  showing a clear association of
the cold gas (at about 20K) with the cooling flow.
This shows that although the AGN provides
a feedback heating, the cooling phenomenon does occur,
with about the expected rate. 

\section{Recent results on cooling flows}
Cooling flows are suspected to occur in the
center of rich galaxy clusters, since the cooling time of the
hot gas is shorter than Hubble time. Until recently, large
flow rates were expected, of the order of 
100 to 1000 M$_\odot$/yr. But the fate of the cooled gas
has always been a mystery, the gas or the stars
formed out of it could not be seen. Attempts to
observe this cold gas either 
in HI (Shostak et al 1983, Dwarakanath et al 1995)
or CO (McNamara et al 1994, Braine \& Dupraz 1994, 
O'Dea et al 1994), have always obtained upper limits.
Only in NGC1275 (Perseus A) has molecular gas been detected 
(Lazareff et al 1989), but being associated to merging galaxies,
its origin was ambiguous. Recently, CO emission has been detected in
3C31 (RXJ0107+32) and 3C264 (A1367) at the center
of cooling flows (Lim et al 2001), and towards
 16 central galaxies of clusters (Edge 2001).

One of the main results of the X-ray satellites
Chandra and XMM-Newton was to change considerably
our view on cooling flows: the deduced cooling rates
have been reduced by at least one order of magnitude;
the old view of quiet and regular, quasi-spherical cooling
has given place to partial and intermittent cooling,
perturbed by re-heating processes, feedback due to the central AGN,
and associated shocks, jets, sound waves, bubbles...
A recent spectacular illustration of this perturbed
cooling is the Chandra image of the cooling flow in Perseus,
with bubbles, gas streaming up and down from the center,
and ripples looking like emitted sound waves (Fabian et al. 2003). 
\index{Perseus}

\section{Detection of CO emission}
We have undertaken a survey of CO emission in 
central galaxies of cooling flow clusters (Salom\'e
\& Combes 2003a), with the IRAM-30m telescope.
The clusters were selected from their previously
derived cooling rates, and their observed H$\alpha$
luminosities. It is indeed expected that the H$\alpha$
emission is also related to the cooling, either 
diretly or because gas
is photoionised by stars formed in the cooling.
Out of 32 galaxies observed, between 6 and 10 detections
were obtained (according to whether both CO(1-0) and CO(2-1)
detected, or a tentative single one). Masses between
3 10$^8$ and 4 10$^{10}$M$_\odot$ are found.
The excitation derived from the ratio of the two
first rotational lines of the CO molecule is compatible
with optically thick cold gas (beam corrected 
ratio close to 1, within the uncertainties).
%
%
\begin{figure}
\centering
\rotatebox{-90}{\includegraphics[height=10.5cm]{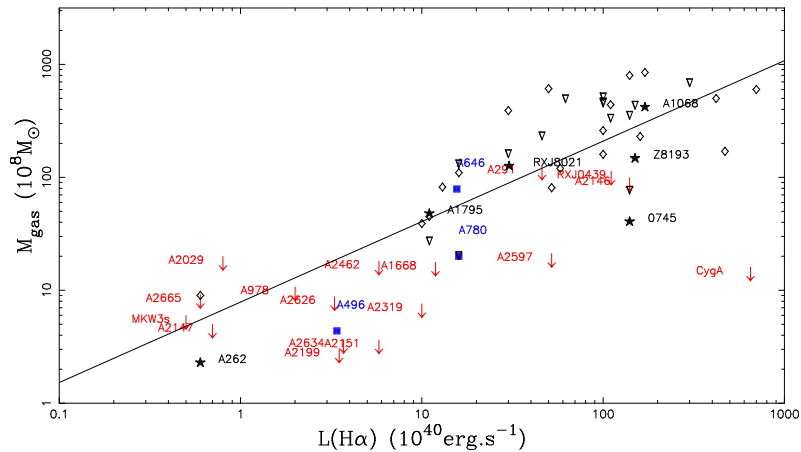}}
\caption{H$_2$ gas mass derived from CO emission, as a
function of H$\alpha$ luminosity; results both from Edge (2001)
and Salom\'e \& Combes (2003a) have been included.}
\end{figure}

Figure 1 reveals a correlation between
the H$_2$ mass detected toward cooling
flows, and their H$\alpha$ luminosity.
This may be interpreted in terms of common
excitation mechanisms, H$\alpha$ is either shocked gas from cooling flows
or gas photo-ionized by the young stars formed out of the cooled gas.

It is also interesting to compare the amount of gas detected
to what is expected from the cooling flow. With respect
to the cooling rates deduced 
 from Einstein (White et al 1997), the CO gas reaches
about 10\% of the expected value.  We have to note that 
the region observed with the IRAM beam is only the center
of the cooling flow. The cooling region extends up to
10 times its radius, and the cooling mass inside a given
radius grows linearly with radius. The H$_2$ gas detected
corresponds to cooling rates during
1 Gyr about 10 times less than the previously computed
value.

Our results do not show any
clear correlation with the radio flux at 1.4 GHz. However,
it is clear that cooling flows help to fuel the central black hole, 
when there is one. About 
71\% of cD galaxies with cooling flow have an AGN, while only 23\%
of cD without cooling flow have one.
It is important to try to derive the gas-to-dust ratio
in the cooled gas, since  dust is expected to be 
destroyed in the hot gas by sputtering. The cooled gas
should be dust depleted. We have derived the dust
mass from IRAS data (assuming a dust temperature
of the order of 35K), and indeed, the dust-to-gas ratio was
small. Cold dust observed at 1mm would yield clearer
results, due to the lower temperature sensitivity,
in the Rayleigh-Jeans domain.

\section{Abell 1795 cooling flow}

The Abell 1795 cluster (z=0.063) has a conspicuous cooling wake at its center
in the form of a long (80kpc) North-South filament (Fabian et al 2001).
The cooling time is about 300 Myr, and the cooling rate
$\sim$ 100 M$_\odot$/yr  within  200kpc (Ettori et al 2002).
The filament presents also H$\alpha$ line emission (Cowie et al 1985),
at the velocity of the cluster.
The central cD galaxy is moving at V=374km/s with respect to the cluster,
and this central oscillation is probably the origin of the wake.

We have mapped CO(1-0) and CO(2-1) emission with the IRAM interferometer
with 3.2" and 1.8" beams respectively. 
The cold gas is clearly coincident with the cooling gas, 
both in X-ray and H$\alpha$, but not with any galaxy
(Salom\'e \& Combes 2003b).  3mm continuum emission is detected at
the position of the AGN, with a flux of 7mJy corresponding to 
a synchrotron spectrum with slope  -0.98. Figure 2
shows the very good coincidence between CO(2-1)
and H$\alpha$ line emission, and its avoidance of the radio jets.
Cold gas may have deflected the expanding radio lobes, 
alternatively
the jet creates a hole (bubble) in the hot gas, which is compressed
at the boundaries, cools down and condenses in molecular gas.

\begin{figure}
\centering
\includegraphics[width=10.5cm]{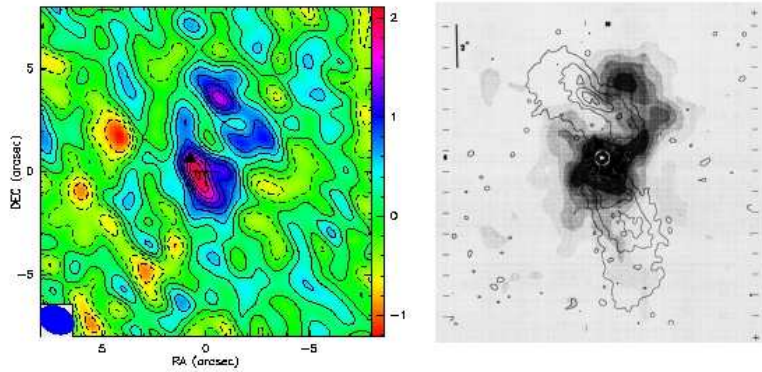}
\caption{{\bf Left} CO(2-1) map obtained with the IRAM interferometer.
{\bf Right} H$\alpha$ +[NII] line emission (grey scale), with 
6cm contours from van Breugel et al. (1984).}
\end{figure}

The interferometer retrieves only 20\% of the 30m flux, 
which means that there exists an important extended CO emission,
in addition to the mapped features. The total H$_2$ mass is 
4.8 10$^9$ M$_\odot$.
The CO kinematics confirm the association of the
molecular gas with the cooling flow.
The CO velocity is not associated to the central galaxy, but to
the cluster (-350km/s), as is the H$\alpha$.

\section{Conclusions}

Out of 32 cooling flow galaxies observed, 6-10 were detected, 
with H$_2$ masses up to 10$^{10}$M$_\odot$, increasing
the total number to 23 (Edge 2001).
The interferometer map  of
Abell 1795 shows CO clearly associated to the cooling wake, and not 
in rotation around the cD galaxy.
The CO emission is  closely associated to H$\alpha$,
confirming the global CO-H$\alpha$ correlation found
with the single dish.
The H$_2$ mass found corresponds to what is expected 
from the cooling rate, 100 M$_\odot$/yr within 200kpc. There
is tight coupling between the AGN and the cold gas.
The AGN creates cavities in the hot gas. Cooling occurs along
the edges of cavities, where CO and H$\alpha$ are observed.

%
\input{referenc}



\printindex
\end{document}

%% file: referenc.tex
%
%

%
%